\documentclass[fleqn,usenatbib]{mnras}

\usepackage{newtxtext,newtxmath}
\usepackage[T1]{fontenc}
\DeclareRobustCommand{\VAN}[3]{#2}
\let\VANthebibliography\thebibliography
\def\thebibliography{\DeclareRobustCommand{\VAN}[3]{##3}\VANthebibliography}
\usepackage{graphicx}	
\usepackage{amsmath}	
\usepackage{color}

\title[Blue GCs as Tracers of Dark Matter - NGC 4278]{A Trail of the Invisible: Blue Globular Clusters Trace the Radial Density Distribution of the Dark Matter - Case Study of NGC~4278}

\author[M. Kluge et al.]{
Matthias Kluge$^{1,2}$\thanks{E-mail: mkluge@usm.lmu.de},
Rhea-Silvia Remus$^{1}$,
Iurii V. Babyk$^{3,4,5}$,
Duncan A. Forbes$^{6}$, \&
Arianna Dolfi$^{6,7}$
\\
$^{1}$University-Observatory, Ludwig-Maximilians-University, Scheinerstrasse 1, D-81679 Munich, Germany\\
$^{2}$Max Planck Institute for Extraterrestrial Physics, Giessenbachstrasse, D-85748 Garching, Germany\\
$^{3}$Center for Astrophysics | Harvard \& Smithsonian, 60 Garden St, Cambridge, MA 02138, USA\\
$^{4}$Department of Physics and Astronomy, University of California, Irvine, CA 92697, USA\\
$^{5}$Main Astronomical Observatory, National Academy of Sciences of Ukraine, 27 Akademika Zabolotnoho St, 03143 Kyiv, Ukraine\\
$^{6}$Centre for Astrophysics \& Supercomputing, Swinburne University, Hawthorn, VIC 3122, Australia\\
$^{7}$ Departamento de Física y Astronomía, Universidad de La Serena, Av. Juan Cisternas 1200 Norte, La Serena, Chile
}

\date{Accepted 2023 March 20. Received 2023 March 13; in original form 2023 January 05}

\pubyear{2023}

\begin{document}
\label{firstpage}
\pagerange{\pageref{firstpage}--\pageref{lastpage}}
\maketitle

\begin{abstract}
We present new, deep optical observations of the early-type galaxy NGC~4278, which is located in a small loose group. We find that the galaxy lacks fine substructure, i.e., it appears relaxed, out to a radius of $\sim$70 kpc. Our $g$- and $i$-band surface brightness profiles are uniform down to our deepest levels of $\sim$28 mag arcsec$^{-2}$. This spans an extremely large radial range of more than 14 half-mass radii. Combined with archival globular cluster (GC) number density maps and a new analysis of the total mass distribution obtained from archival Chandra X-ray data, we find that the red GC subpopulation traces well the stellar mass density profile from 2.4 out to even 14 half-mass radii, while the blue GC subpopulation traces the total mass density profile of the galaxy over a large radial range. Our results reinforce the scenario that red GCs form mostly in-situ along with the stellar component of the galaxy, while the blue GCs are more closely aligned with the total mass distribution in the halo and were accreted along with halo matter. We conclude that for galaxies where the X-ray emission from the hot halo is too faint to be properly observable and as such is not available to measure the dark matter profile, the blue GC population can be used to trace this dark matter component out to large radii.

\end{abstract}

\begin{keywords}
galaxies: star clusters: general --- galaxies: haloes --- galaxies: structure --- galaxies: photometry
\end{keywords}

\section{Introduction} \label{sec:intro}

Globular Clusters (GCs) are among the oldest objects in our Universe and are found around most (large) galaxies. Generally, it has been known for a while that the more massive a galaxy, the more numerous its population of globular clusters (GCs) \citep{blakeslee1997}. More recently, it has been found that the total number (or mass) of GCs in a given galaxy is actually an excellent tracer of the total halo mass of the galaxy, over many orders of galaxy luminosity and galaxy type \citep{Spitler2009,Harris2015}, even down to the dwarf galaxy regime \citep{forbes2018c,Burkert2020}. This indicates that there is a tight correlation between the most dominant part of the galaxy, namely its dark matter content, and its GC population. This is supported by the E-MOSAIC model that showed that the 2D spatial distribution of GCs in high mass galaxies closely follows the distribution of total mass \citep{Reina-Campos2022a}. However, the origin of GCs is still subject to debate \citep{Forbes2018}, and thus also the connection between a galaxies formation and GC population is under investigation.

It is also well known that the GCs around massive galaxies can be divided into subpopulations of different metallicities: red metal-rich GCs and blue metal-poor GCs \citep[e.g.,][]{Usher2012}. For most galaxies, the blue subpopulation is radially more extended than the red subpopulation \citep[e.g.,][]{peng2004,Brodie2006,schuberth2010,Strader2011}. In addition, \citet{Harris2015} showed that the scaling relation between the total mass and the number of GCs is closer to linear for blue GCs than for red GCs. This gave rise to the idea that red GCs are formed in-situ in the galaxies, while blue GCs are mostly formed early in mass-poor galaxies and thus mostly accreted onto the galaxy through merging processes. As such, they are mostly ex-situ in origin \citep[e.g.,][]{2014ApJ...796...10L,Forbes2018b}. Semi-analytical and semi-empirical models have supported this idea, showing that the observed correlations are a combination of such in-situ and accretion-driven GC formation processes \citep[e.g.,][]{beasley2002,Valenzuela2021,Reina-Campos2022b}.

If the origin of the red and blue GCs in a galaxy is indeed of in-situ and accreted nature, respectively, then they could also be tracing different parts of the galaxy. In fact, in some galaxies, albeit not all, blue and red GCs are found to exhibit very different radial number density distributions \citep[e.g.,][]{pota2013}. \citet{Forbes2012} showed that the projected density profile of the red GCs follows the stellar distribution while the blue GC density profile was a better match to the projected density of the hot halo gas as traced by the X-ray emission in large elliptical galaxies. Similar results had already been found for the case of the cD galaxy NGC~1399 by \citet{Forte2005}. More recently, \citet{dolfi2021} combined planetary nebulae, GCs, and stellar light to examine the galaxy internal kinematics out to 5-6 half-light radii for 9 SLUGGS galaxies, finding that the red GCs indeed trace the light out to large radii. However, no X-ray profiles were included in that work.

In this study we focus on the early-type galaxy NGC~4278 and its GC system. This galaxy was chosen particularly due to the fact that the radial density profiles of the red and blue GCs show very different slopes \citep{pota2013}, and deep Chandra X-ray archival data are available. Unlike in previous studies by \citet{Forbes2012} and \citet{Forte2005}, this galaxy is not a massive cluster or group central galaxy, but rather lies in the low-density environment of the Coma I group, therefore is comparably low in total mass. Thus, it extends the previous studies down to the lower mass end, where X-ray data are increasingly difficult to obtain. It has an old stellar population with no sign of ongoing star formation, but it does host a weak radio AGN \citep[e.g.][]{pellegrini2012}, shows signs of outflows in the central region \citep{bodgan2012}, and contains a large-scale HI disk \citep[see][for further details]{Usher2013}. It has a lower mass companion galaxy, NGC~4283, approximately 16\,kpc away in projection at an assumed distance of 15.4 Mpc. NGC~4278 is part of the SLUGGS survey which examined in detail the GC systems of two dozen nearby early-type galaxies \citep{Brodie2014}. HST imaging of the NGC~4278 GC system was presented by \citet{Usher2013}. They estimated a total number of 1378 GCs and a GC specific frequency of S$_N$ = 6, which suggests a fairly rich GC system for its host galaxy. \citet{pota2013} examined the kinematics of the blue and red GCs while \citet{foster2016} used the stellar kinematics to examine 2D kinematic maps out to 2 half-light radii. These studies have found NGC~4278 to be a rather undisturbed early-type galaxy, with no indications of interactions between NCG~4278 and its companion galaxy NCG~4283, neither in the kinematics nor in the light distribution or the GC population. Despite a kinematically distinct fast-rotating core at its very center, it is a slow-rotating galaxy out to large radii. However, using low-mass X-ray binaries as tracers, \citet{dabrusco2014} found evidence of previous accretion of dwarf galaxies that have been disrupted in the past.

In this work, we combine multi-messenger information to study the origin of the GC subpopulations in NGC~4278. Therefore, we obtain an extremely deep radial stellar light profile and compare the distribution of stellar and total mass to that of the blue and red GC subpopulations. Throughout the paper, we assume a distance to NGC 4278 of 15.4\,Mpc \citep{Tully2013} and a physical scale of 0.074\,kpc\,arcsec$^{-1}$.

\begin{figure*}
    \includegraphics[width=\linewidth]{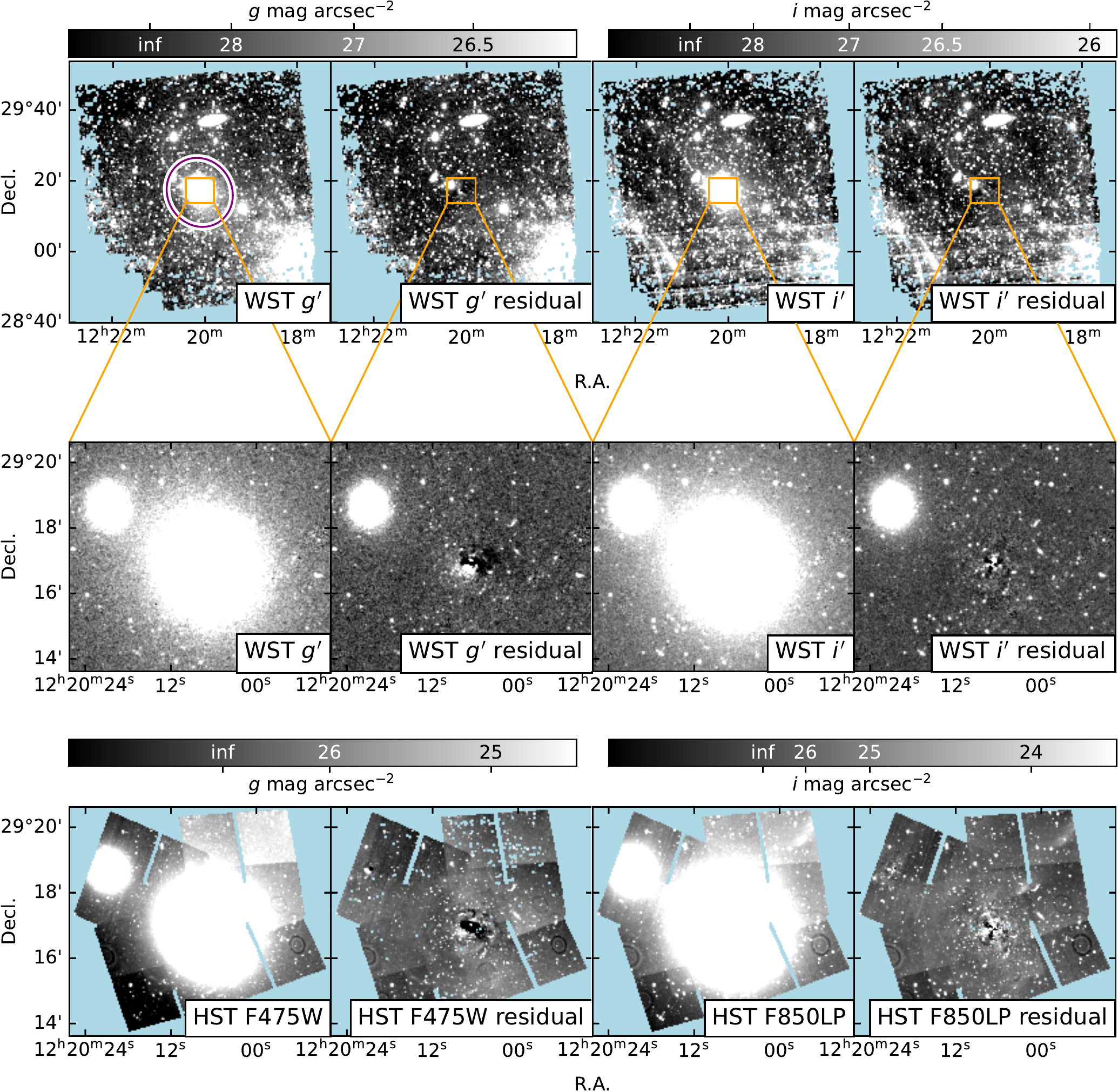}
    \caption{Observations of the stellar light using the Wendelstein (WST) 43\,cm telescope (top two rows) and archival HST observations (bottom row). The surface brightness scale is shown above the first and third rows. The second row shows a zoom of the inner regions as indicated by the orange lines. The original images are shown in the first and third columns, whereas the residuals after subtracting isophote models (and gradient for the HST data) are shown in the second and fourth columns. For the HST data, we have subtracted isophote models for both galaxies: the target galaxy NGC 4278 (centered) and the companion galaxy NGC 4283, which is visible in the top left of the second and third rows. \label{fig:images}}
\end{figure*}

\section{Observations of the Stellar Light}

Imaging observations were carried out in the $g'$ and $i'$ bands with the 43\,cm telescope at the Wendelstein Observatory. Although the collecting area is small, the setup has been shown to be well suited for deep imaging studies \citep{Kluge2020}. The camera has a field of view of $43\arcmin\times43\arcmin$, which is covered by one single CCD with a pixel scale of 0.634\arcsec\,pixel$^{-1}$. The basic data reduction is similar to the procedure described in \cite{Kluge2020}. One exception is that we only mask bright foreground stars instead of subtracting models from them. This is sufficient because the apparent galaxy size is large compared to the PSF of a typical bright star (see Figure \ref{fig:images}). However, the large extent of NGC 4278 complicates the background subtraction, because it covers a significant fraction of the field of view. We thus perform the background subtraction differently using offset sky exposures.

\begin{figure*}
    \centering
    \includegraphics[width=\linewidth]{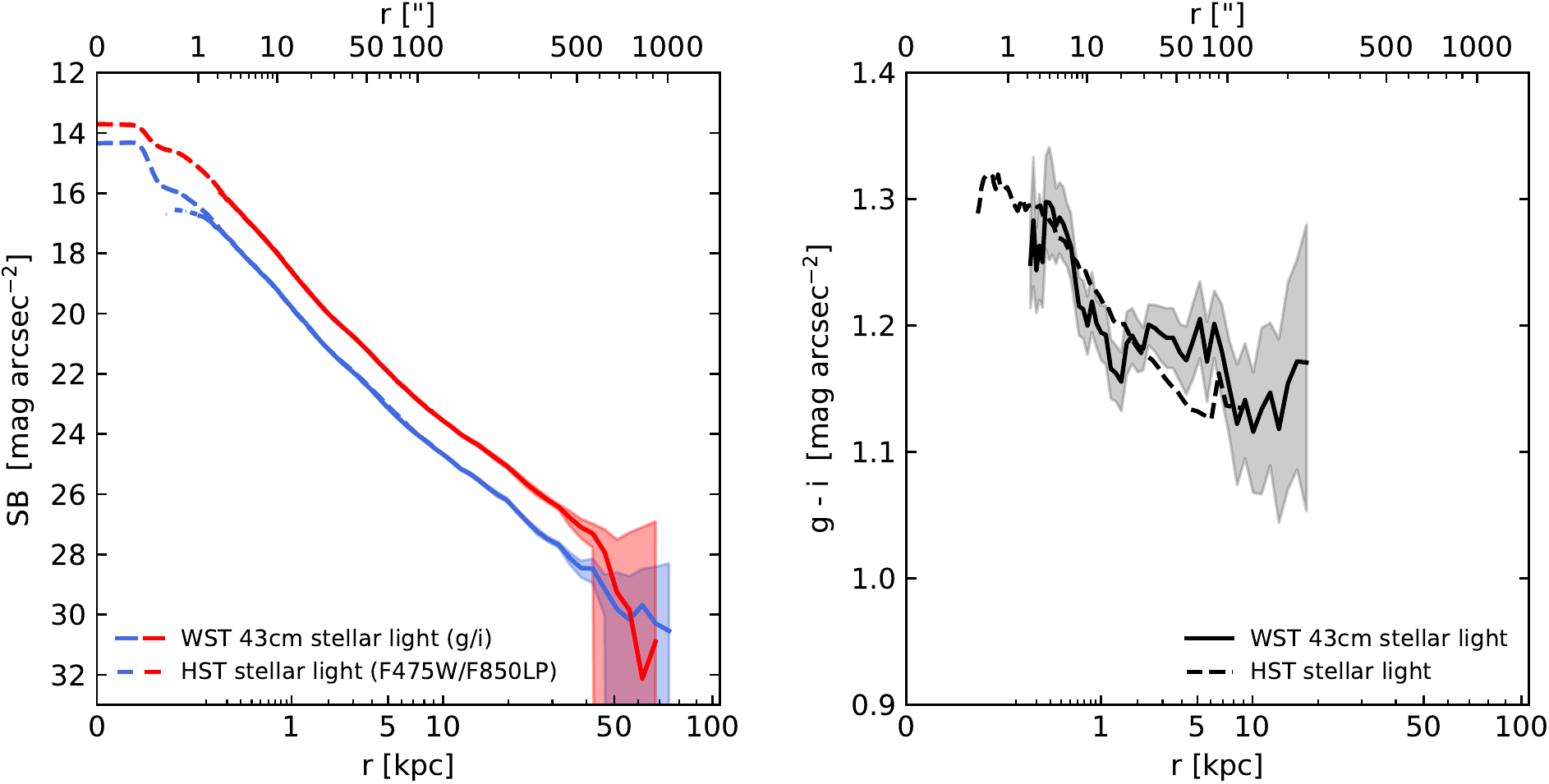}
    \caption{Surface brightness (SB) profiles (left panel) and $g-i$ color profiles (right panel) of the stellar light. The radius $r$ is measured along the effective axis $r=\sqrt{ab}$, where $a$ ($b$) is the semi-major (semi-minor) axis radius. The x-axis is scaled with a constant step size in $r^{1/4}$. Data from deeper WST data is shown using continuous lines and higher resolution HST data is shown using dashed lines. The shades correspond to the 1$\sigma$ uncertainties. The WST profiles are reliable down to ${\rm SB}\sim 28$ mag arcsec$^{-2}$. The galaxy becomes slightly bluer with radius. \label{fig:lightprofiles}}
\end{figure*}

NGC 4278 is located at R.A. = 12:20:07 and Decl. = +29:16:51 (J2000). In addition to this galaxy pointing, we have observed a 2\degr~offset sky pointing at R.A. = 12:18:48 and Decl. = +31:15:16. This region is devoid of bright sources and thus allows us to subtract the background in the galaxy pointing. Galaxy and sky pointings were alternated during the observations. The exposures of each pointing were repeatedly performed using a spiral-shaped dither pattern with 13 steps and 5\arcmin~step size. The two sky pointings ($S_1$ and $S_2$) that bracket a galaxy exposure are combined and subtracted from that particular galaxy exposure. Prior to combining the two sky images, their flux levels are matched to reduce biases due to a timely varying sky brightness:

\begin{equation}
    S_{i,{\rm scaled}} = \frac{S_i}{{\rm mean}(S_i)}  \cdot \frac{{\rm mean}(S_1) + {\rm mean}(S_2)}{2}.
\end{equation}

This flux matching is done for both $S_i=S_1$ and $S_i=S_2$. The mean pixel value is calculated only for pixels, which are unmasked in \textit{both} sky images. Afterward, the rescaled sky images are averaged and subtracted from the bracketed galaxy exposure. Regions, which are masked in both sky images remain masked in the sky-subtracted galaxy image too. These sky-subtracted galaxy images are finally resampled and co-added. Panels 1 and 3 in the top row of Figure \ref{fig:images} show the results. Zoom-ins on the galaxy center are shown in the middle row. The isophote with a 10\arcmin~semi-major axis radius is overplotted onto the top left panel by the purple-white ellipse. The surface brightness profile is reliable up to this radius (see Figure \ref{fig:lightprofiles}, left panel). It shows that the stray light in the southwest is sufficiently far away. Moreover, we emphasize that any visible stray light is masked before the SB profile is measured.

One or more bright stars outside the field of view produce curved, large-scale stray light in the southeast region of NGC 4278. Because of strong contamination, we discard 4/13 full dither steps in that region for the $g'$ band and 2/13 full dither steps for the $i'$ band. The final integration time of the galaxy pointing totals $52\times3{\rm\,min}=2.6{\rm\,hr}$ in the $g'$ band and $145\times3{\rm\,min}=7.25{\rm\,hr}$ in the $i'$ band.

The surface brightness (SB) profiles are measured following the procedure described in \cite{Kluge2020} with the modifications described in Kluge \& Bender (in prep.). In brief, we mask dust, straylight, and all sources besides the galaxy and use the {\tt python} package {\tt photutils} \citep{Bradley2021} to fit ellipses to the isophotes in the $i'$ band. 

They are fixed beyond $\sim$60\arcsec~semi-major axis radius to avoid distortions from the imperfectly masked stellar halo of the companion galaxy NGC 4283. The SB profile is measured using the median flux in annuli around these ellipses. In order to measure color gradients, we use the same ellipses for both $g'$ and $i'$ bands.

To quantify the impact of the residual stray light, especially in the southwest, we coadd all exposures of the non-discarded dither steps separately. The SB profile of NGC 4278 is then measured on those stacks independently. The mean profile is shown by the continuous lines in Figure \ref{fig:lightprofiles}, left panel. Error shades correspond to the 1$\sigma$ scatter. The image residuals after subtracting isophote models are shown in Figure \ref{fig:images}, top and middle rows, panels 2 and 4.

In the center, the SB and color profiles are replaced by our measurements on archival \textit{Hubble Space Telescope} (HST) ACS/WFC images (dashed lines). This has the advantage of increasing the central resolution from the Wendelstein PSF FWHM of 3.4\arcsec~and 3.85\arcsec~in the $g'$ and $i'$ band, respectively, to $\sim$0.08\arcsec. A higher resolution improves the inner SB profile but also allows more precise masking of the central dust. The HST image mosaics are shown in Figure \ref{fig:images}, bottom row, panels 1 and 3. Furthermore, we measure and subtract models of both NGC 4278 and 4283 to look for unrelaxed accretion signatures and tidal interactions between the two galaxies. Apart from the central dust in NGC 4278, the galaxies appear relaxed without any fine structure (panels 2 and 4).

The HST and Wendelstein SB profiles are merged at $r=1.77$\,kpc, where both color profiles cross (see Figure \ref{fig:lightprofiles}, right panel). This is outside the region of the PSF and dust contamination, but inside the region where the contamination by the companion galaxy is not yet significant. 

Photometric zero points are calibrated to the SDSS $g$ and $i$ filter systems. For this reason, we neglect the $'$ from here on when referring to the $g'$ and $i'$ WST filters. We measure the SB profiles on archival SDSS images and shift the Wendelstein SB profiles until they match with the SDSS SB profiles. Galactic dust extinctions $A_g=0.098$ and $A_i=0.051$ are applied using the maps from \cite{Schlafly2011}.

Stellar masses are calculated using the color-dependent mass-to-light ratio $\Psi$ in the $g$ band following \cite{Roediger2015}:

\begin{equation}
    \log(\Psi)_g = 1.379 (g - i) - 1.067.
\end{equation}

We fix the color inside $r<1$\,kpc to $g-i=1.23$ because of the strong dust absorption and outside $r>13$\,kpc to $g-i=1.13$ because of the high uncertainties.

Beyond $r>1.77$\,kpc, we convert the $g$-band SB profile to the stellar mass profile and inside that radius, the $i$-band SB profile is used. The $g$-band profile has the advantage of a more stable background, while the $i$-band profile is less sensitive to dust absorption.

By fitting a single S\'ersic function \citep{Sersic1968} to the stellar mass profile, we obtain an effective radius of $r_{\rm e} = 2.875\pm0.076$\,kpc and S\'ersic index $n=5.96\pm0.16$ (see Table \ref{tab:fitparams}). 
Although the S\'ersic index is consistent with \cite{forbes2017b} ($n=6.2\pm0.13$, $r_{\rm e}=2.09\pm0.013$\,kpc), our measured effective radius is larger. This is likely due to intrinsic deviations from a perfect S\'ersic profile, a smaller fitting range $r<100\arcsec$ used by \cite{forbes2017b} and the blue color gradient (see Figure \ref{fig:lightprofiles}, right panel), which makes NGC 4278 appear more compact in the 3.6$\mu$m infrared Spitzer imaging data used by \cite{forbes2017b}.

By integrating the best-fit S\'ersic function to the infinite radius, while respecting the radially varying ellipticity, we obtain a total stellar mass $\log(M_* [\rm{M}_{\odot}]) = 10.913\pm0.006$. The statistical uncertainty from the fit likely underestimates the real uncertainty because of intrinsic deviations from a perfect S\'ersic profile and uncertainties in the stellar mass-to-light ratio. Nevertheless, our value is consistent with \cite{forbes2017b}, who measured $\log(M_* [\rm{M}_{\odot}]) = 10.95\pm0.1$.

\section{Total Mass Profile} \label{sec:totalmass}

\begin{figure}
	\includegraphics[width=\linewidth]{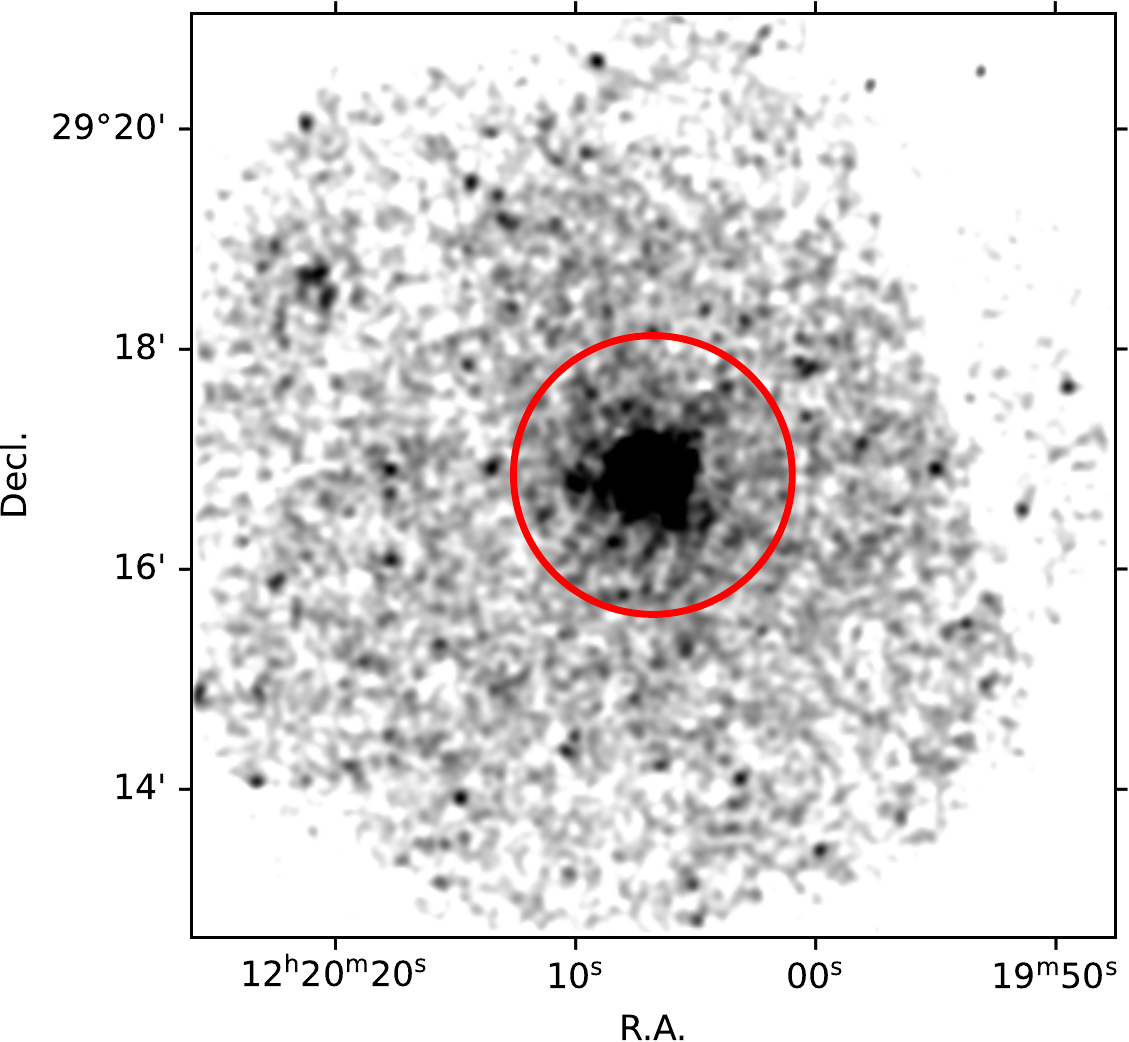}
	\caption{Co-added Chandra X-ray image of NGC 4278. For better visibility, bright sources are masked and the image is smoothed with a Gaussian kernel with 5 pixel standard deviation. The red circle has a radius of 5.6\,kpc, which is the maximum radius of the derived total mass profile using a beta model. \label{fig:chandra}}
\end{figure}

\begin{figure}
	\includegraphics[width=\linewidth]{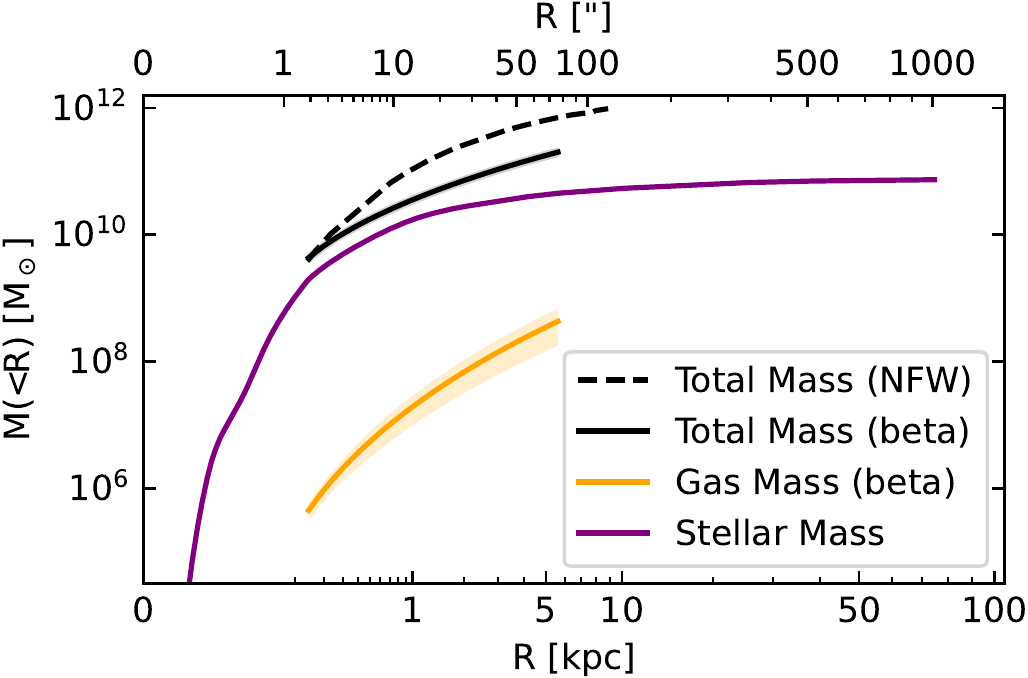}
    \caption{Curves of growth of the 3-D total mass (black), gas mass (yellow), and stellar mass (purple) profiles. Total mass profiles are shown by the dashed line for the NFW model and by the continuous line for the $\beta$ model. \label{fig:mass_profiles}}
\end{figure}

To constrain the total mass profile of NGC 4278, we use 161 hours of archival Chandra X-ray imaging data. The coadded image is shown in Figure \ref{fig:chandra}. We have masked bright sources before smoothing the image for better visibility. The X-ray emission from discrete sources accounts for $\sim3\%$ of the total X-ray emission of the hot gas assuming a multi-component spectral model. The red circle marks the largest radius $r=5.6$\,kpc, for which the cumulative total mass profile $M(<R)$ is constrained (see Figure \ref{fig:mass_profiles}). The procedure is fully described in \cite{Babyk2018}. In brief, we fit a $\beta$ model \citep{Cavaliere1978} ($\beta=0.58\pm0.01$) to the X-ray surface brightness profile. We then calculate the cumulative gas and total mass profiles by assuming spherical symmetry and hydrostatic equilibrium. To estimate the model uncertainty, we repeat the procedure assuming that the total mass follows an NFW profile \citep{Navarro1996}.

The results are shown in Figure \ref{fig:mass_profiles} by the black lines. In purple, we overplot the cumulative 3-D stellar mass profile. Therefore, we convert the 2-D surface mass density $\Sigma(r)$ to a 3-D mass density $\rho(R)$ by assuming spherical symmetry \citep{Binney2008}:
\begin{equation}
    \rho(R) = -\frac{1}{\pi} \int_R^\infty \frac{\rm{d}r}{\sqrt{r^2 - R^2}} \frac{\rm{d}\Sigma}{{\rm d}r}
\end{equation}
and integrate it in shells to obtain the cumulative stellar mass profile.

To directly compare the 2-D stellar surface mass profile, the 2-D GC number density profiles, and the total surface mass profile (see Section \ref{sec:comparison}), we convert the cumulative total mass $M(<R)$ to a radial 3-D mass density $\rho(R)$ by assuming spherical symmetry

\begin{equation}
    \rho(R) = \frac{1}{4\pi R^2} \frac{{\rm d}M(R)}{{\rm d}R}.
\end{equation}

Then, we project the 3-D mass density to a 2-D surface mass density $\Sigma(r)$ at a projected radius $r$ via an Abel transform \citep{Binney2008}:

\begin{equation}
    \Sigma(r) = 2 \int_r^\infty \frac{\rho(R) R}{\sqrt{R^2 - r^2}} {\rm d}R.
\end{equation}

The projected circular velocity is given by

\begin{equation}
    v_{\rm circ} = \frac{G M(<R)}{3R}, \label{eq:vcirc}
\end{equation}

where G is the gravitational constant and $R$ is the 3-D radius. The $\beta$ model is isothermal outside of the very center. Hence, the circular velocity is constant in this case at $v_{\rm circ}=226\pm13$\,km\,s$^{-1}$. For the NFW model, the circular velocity varies with radius (see Section \ref{sec:gckin}).

\begin{figure*}
	\includegraphics[width=\linewidth]{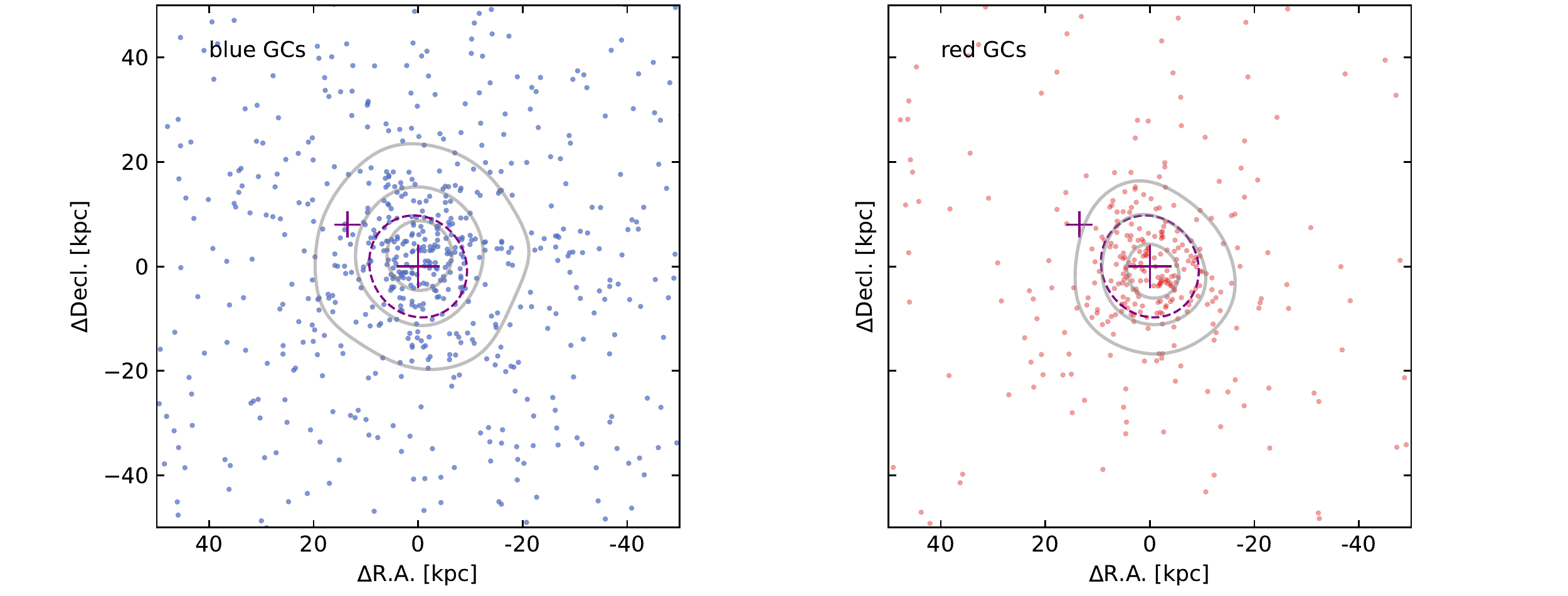}
	\caption{Spatial distribution for the spatially complete photometric Subaru sample of the blue (left) and red (right) GCs. The grey contours mark constant values of the kernel density estimates. The large purple cross indicates the center of NGC 4278 and the small cross indicates the center of the companion galaxy NGC 4283. For comparison, an isophote of the stellar light is overplotted by the dashed purple ellipse. North is up, east is left. \label{fig:gc_distribution_sub}}
\end{figure*}

\begin{figure}
	\includegraphics[width=\linewidth]{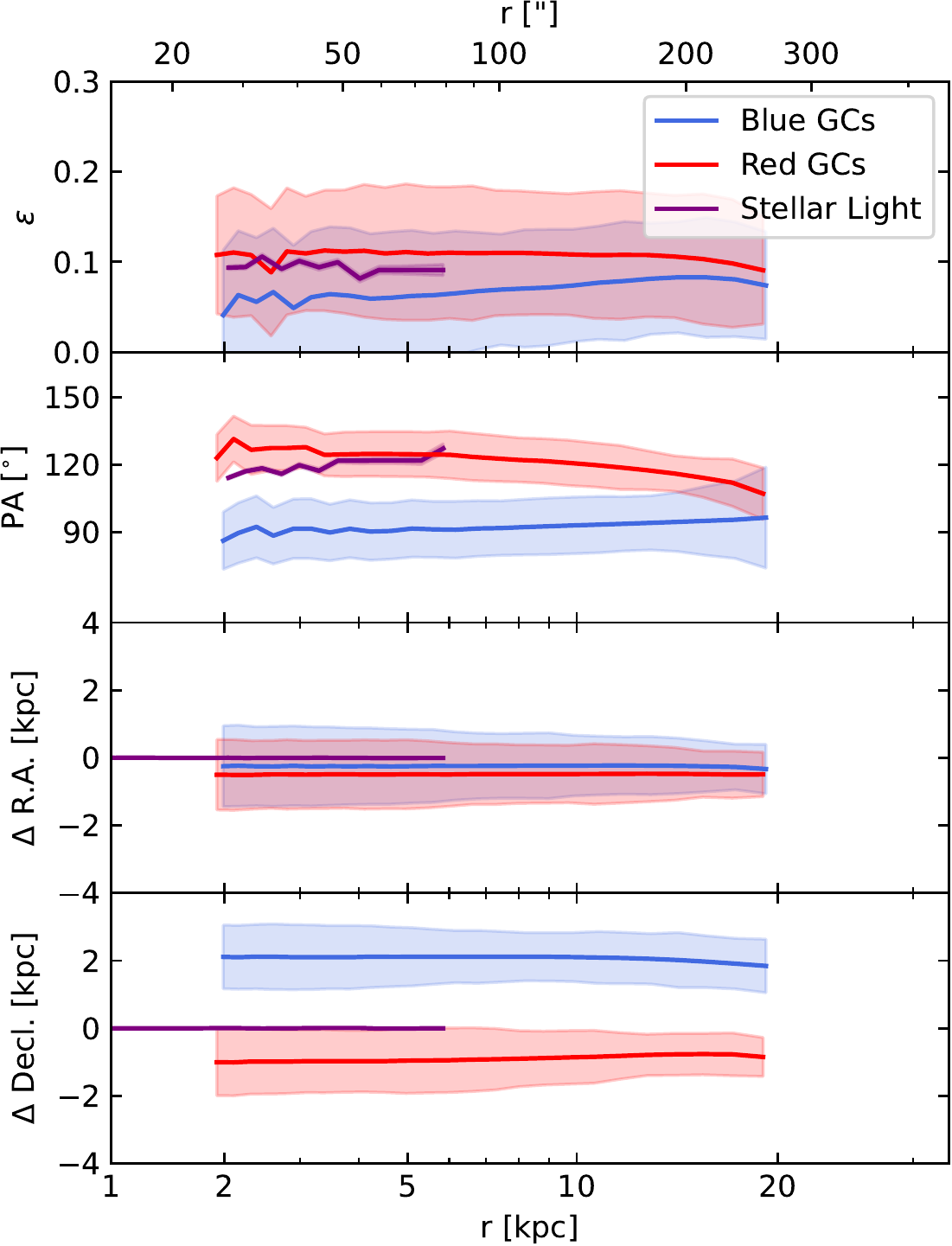}
	\caption{Photometric properties of the Subaru sample for the blue and red GCs and the stellar light (black). From top to bottom: ellipticity, position angle, offset in right ascension, and offset in declination. Note that isophotal shape profiles are less stable and do not extend as far out as the SB profiles. This is mostly because the stellar light of NGC 4278 strongly overlaps in projection with that of NGC 4283 beyond $r\sim10$\,kpc. Also, the galaxy is almost round and so the position angle is poorly defined.
	\label{fig:gc_isophotes}}
\end{figure}

\begin{figure*}
	\includegraphics[width=\linewidth]{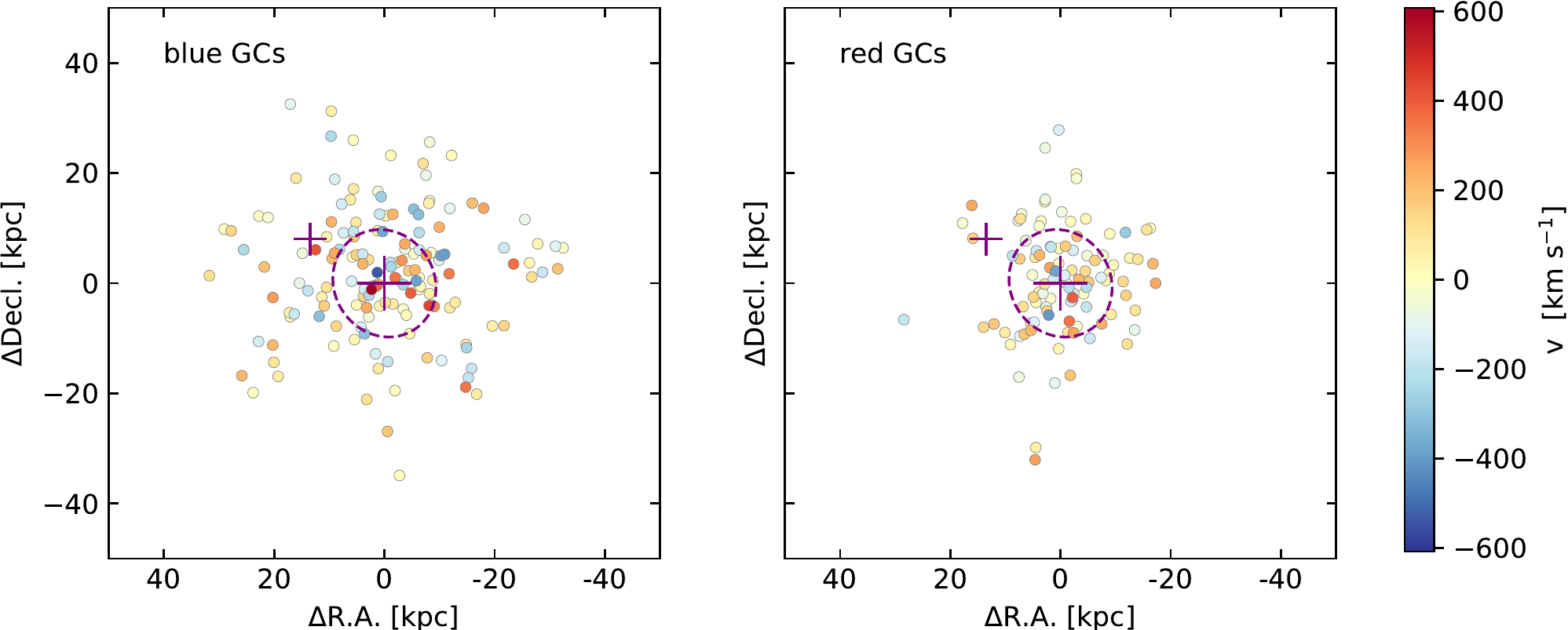}
	\caption{Spatial distribution for the spectroscopic sample of the blue (left) and red (right) GCs color-coded by their line-of-sight velocity. The large purple cross indicates the center of NGC 4278 and the small cross indicates the center of the companion galaxy NGC 4283. For comparison, an isophote of the stellar light is overplotted by the dashed purple ellipse. North is up, east is left. \label{fig:gc_distribution}}
\end{figure*}

\begin{figure}
    \includegraphics[width=\linewidth]{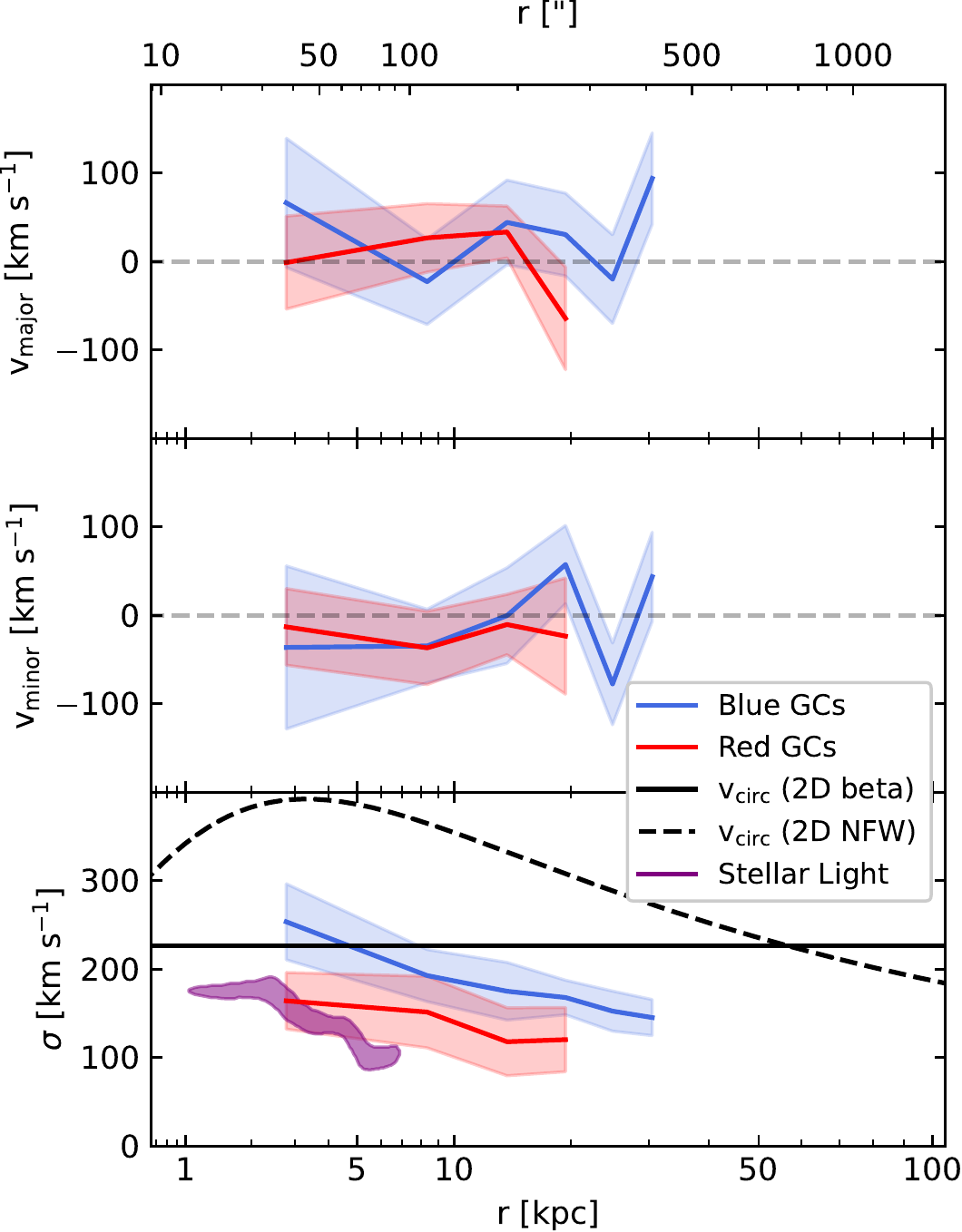}
    \caption{Kinematic properties of the blue and red GCs as well as the stellar light (purple) and total mass (black). From top to bottom: rotation velocity along the major axis, minor axis, and velocity dispersion. The stellar velocity dispersion profile is taken from \citet{foster2016}. The purple shade covers the 2$\sigma$ confidence interval. The circular velocity $v_{\rm circ}$ is calculated with Equation (\ref{eq:vcirc}). \label{fig:gc_v_sigma}}
\end{figure}

\section{Globular Cluster Spatial Distribution and Kinematics}

\subsection{GC Spatial Distribution} \label{sec:gcspatial}

GC number density profiles are taken from \cite{Usher2013} (see also \citealt{pota2013}) for both HST ACS and Subaru Suprime-Cam data sets. This sample is selected photometrically without spectroscopic confirmation. Hence, it extends to larger radii, but it also suffers from some contamination by other sources (see Section \ref{sec:redtracestellarmass}). The limiting brightness is $z=23.44$\,mag, which is one magnitude fainter than the turnover magnitude of the GC luminosity function \citep{Usher2013}. The colors, which are used to split the GC candidates into the blue and red subpopulations are $V-I=1.05$\,mag for the Subaru data and ${\rm F475W}-{\rm F850LP}=1.08$\,mag for the HST data.

We show in Figure \ref{fig:gc_distribution_sub} that blue (left panel) and red (right panel) GCs have different spatial distributions. While the red GCs are more concentrated around NGC 4278, the blue GCs extend to larger galactocentric radii. That impression is confirmed by the shallower number density profile of the blue GCs (see Section \ref{sec:comparison}).

Both, the blue and red GCs are distributed relatively spherical around NGC 4278. We visualize this in Figure \ref{fig:gc_distribution_sub} using the grey contours, which are measured on Gaussian kernel density estimates of the GC distributions. For comparison, an elliptical isophote of NGC 4278 is shown by the purple dashed contour. More quantitatively, Figure \ref{fig:gc_isophotes} compares the isophotal shapes of the stellar light and the GCs. The ellipticities, position angles, and spatial offsets of the GCs are measured by fitting ellipses using the tool \texttt{photutils} \citep{Bradley2021} to the kernel density estimates of the GC distributions. Uncertainties are estimated using 100 bootstrap realizations of the GCs.

The ellipticity profile in the top panel confirms the relatively round shape of the GCs distribution, consistent with the stellar light. The position angle (second panel) of the red GCs agrees well with the stellar light, whereas the blue GCs are rotated by $30\degr$ with a $2\sigma$ significance. However, no strong conclusion can be made from this, because the ellipticity is small. The two bottom panels show spatial offsets with respect to the center of NGC 4278. We find that the GCs are well centered around NGC 4278 with the blue GCs being marginally offset toward the north by $2\pm1$\,kpc.

\subsection{GC Kinematics} \label{sec:gckin}

The catalog of spectroscopically confirmed GCs is taken from \cite{forbes2017c}. It is not complete in a spatial sense nor is it as deep in terms of magnitude ($z < 21$\,mag) compared to the photometrically selected sample in Section \ref{sec:gcspatial}. However, it is free from contamination by other sources. Following \cite{pota2013}, we discard three GCs whose positions and radial velocities are consistent with the neighboring galaxy NGC 4283. We find no overdensity in the region around NGC 4283 (see Figure \ref{fig:gc_distribution}) and thus conclude that the GC kinematics are unbiased.

We measure the GC line-of-sight velocity dispersion at the location of each GC by the standard deviation of the line-of-sight velocities of the ten nearest neighbors including the GC itself. To calculate the radial profile in Figure \ref{fig:gc_v_sigma}, third panel, we average the velocity dispersions in 10\,kpc wide circular annuli. Reducing the number of neighbors to five does not change the dispersion profile significantly, but it increases the scatter.

Rotation curves are measured with respect to the projected major and minor axes (Figure \ref{fig:gc_v_sigma}, top and middle panels). Major-axis rotation corresponds to a rotation signal along the major axis, that is, rotation around the minor axis, and vice versa. Minor axis rotation indicates prolate objects, whereas major axis rotation indicates oblate objects. The rotation curves are obtained by fitting the amplitude of a cosine function to the GC line-of-sight velocities against their azimuthal angles with respect to the major or minor axis. For each radial bin, the GCs are selected in the same circular annuli that are used to calculate the velocity dispersion profiles.

The GC velocity dispersion is smaller for the red GCs than for the blue GCs at all measured radii. This is consistent with the more compact spatial distribution of the red GCs (see Section \ref{sec:gcspatial}).

\begin{figure*}
	\includegraphics[width=0.99\linewidth]{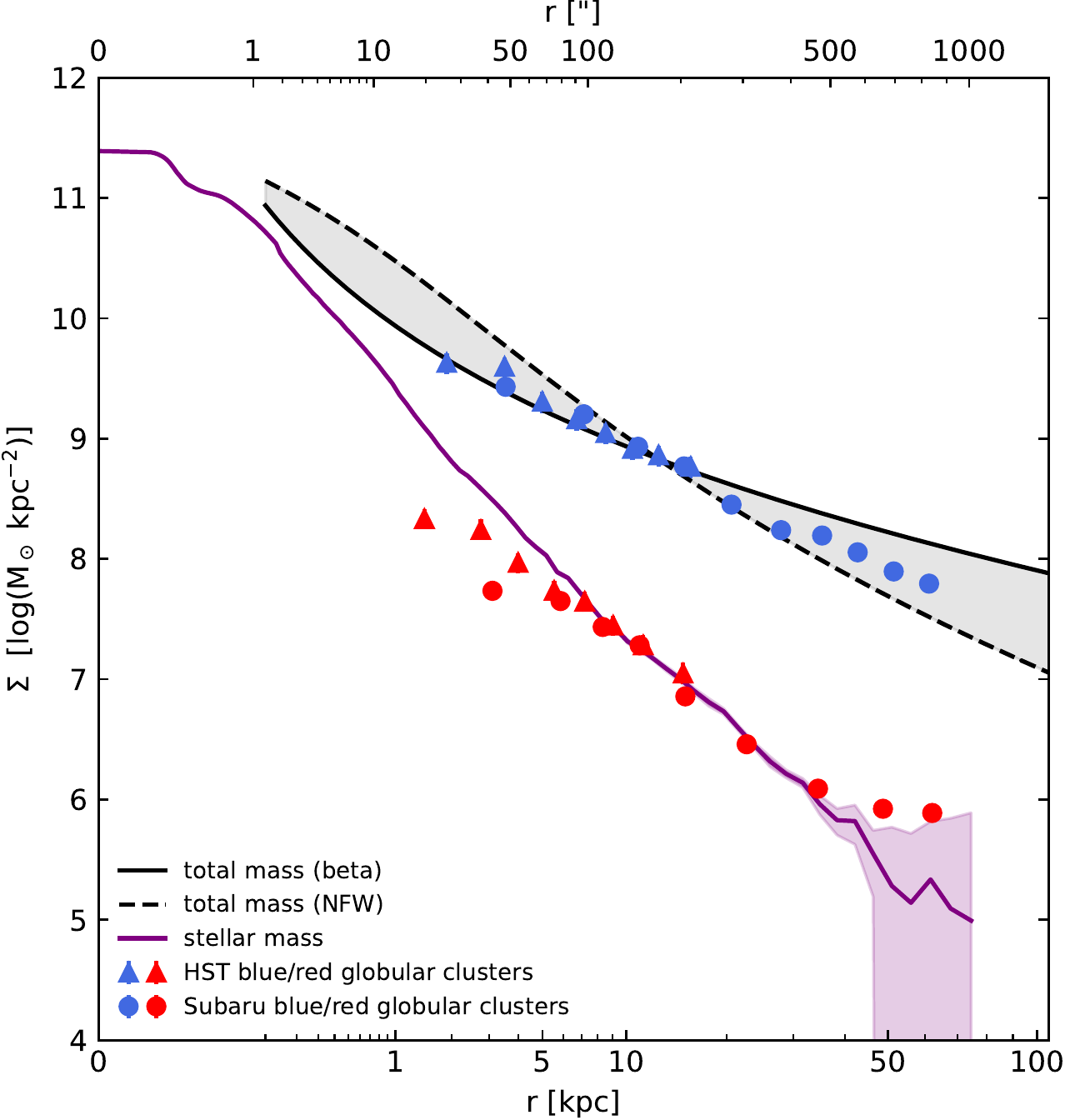}
    \caption{Profiles of stellar mass (purple), total mass (black), red GC number density (red), and blue GC number density (blue). Shades correspond to 1$\sigma$ uncertainties for the stellar mass profile and model uncertainties for the total mass profile. For the GC profiles, the uncertainties are given by the error bars, which are generally smaller than the symbol sizes. The red GC profile is shifted vertically to match the stellar mass profile in the region $7<r<40$\,kpc. The blue GC profile is shifted vertically to match the total mass profile. \label{fig:massprofiles}}
\end{figure*}

\section{Stellar Kinematics} \label{sec:stellarkinematics}

Data of the stellar kinematics of NGC 4278 are taken from \cite{foster2016}. They were measured with the Keck/DEIMOS multi-slit spectrograph as part of the SLUGGS survey \citep{Brodie2014}. Figure A1 in \cite{foster2016} shows the central major-axis rotation with a velocity of $v\approx50$\,km\,s$^{-1}$ inside $r<2$\,kpc. This rotation is aligned with the presence of dust (see Figure \ref{fig:images}), which may indicate a past wet merger. At large radii, there is no significant rotation, consistent with the lack of rotation in the GC system (see Figure \ref{fig:gc_v_sigma}, top and middle panels). The bottom panel in Figure \ref{fig:gc_v_sigma} shows that the central stellar velocity dispersion is consistent with that of the red GCs, but the profile decreases more steeply with radius.

\begin{figure*}
    \includegraphics[width=\linewidth]{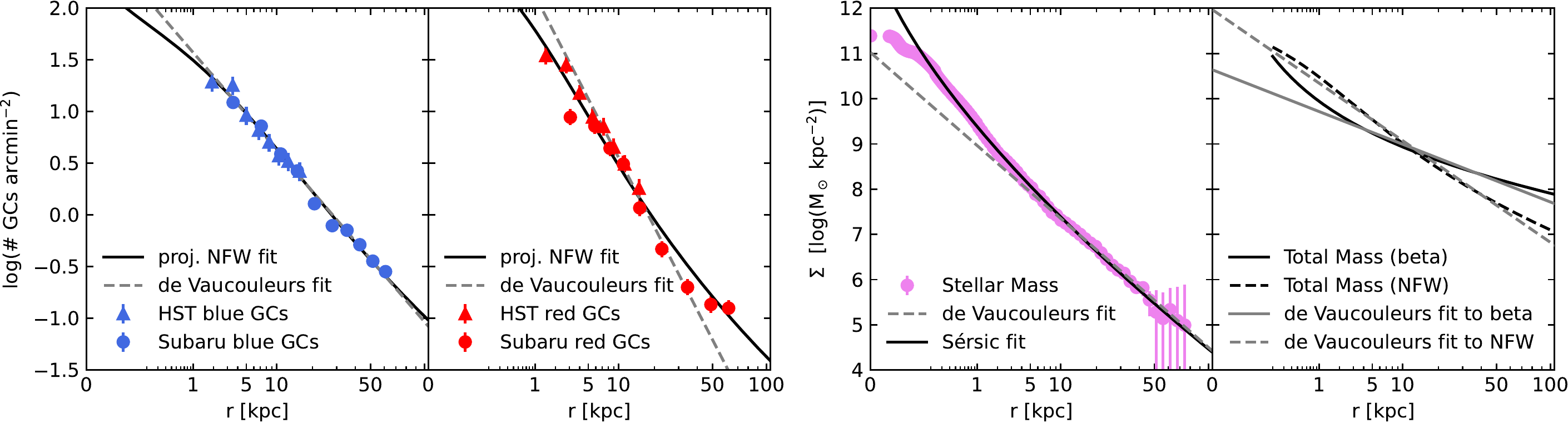}
    \caption{Fits the GC number density and mass profiles. Best-fit parameters are given in Table \ref{tab:fitparams}. For the blue GCs, the projected NFW profile provides a good fit. This is not the case for the red GCs, where the projected NFW profile is too shallow. A S\'ersic fit to the stellar mass profile provides a reasonable fit. There are small, intrinsic variations in the surface mass profile, which cannot be captured by a single S\'ersic function. For the total mass profiles, both the $\beta$ and NFW models are shown. De Vaucouleurs profile fits are shown by grey lines. \label{fig:fits}}
\end{figure*}

\section{Combining Multi-Messenger Data} \label{sec:comparison}

In this section, we compare the spatial and kinematic properties of the four analyzed components of NGC 4278: stellar mass, total mass, red GCs, and blue GCs. Figure \ref{fig:massprofiles} presents the main result of this work: \textit{the red GCs trace the stellar mass while the blue GCs trace the total mass of NGC 4278}. The red GC number density profile is steeper than the blue one. Similarly, the stellar mass density profile (purple) is steeper than the total mass density profile (black). To demonstrate the similarity of the profiles, we shift the blue and red GC profiles along the y-axis until they match the stellar mass and total mass profiles, respectively. After the shift, we quantify the agreement using a reduced $\chi^2$ statistic:

\begin{equation}
    {\rm red.~} \chi^2 = \sum_i \frac{ (\log n_{\rm GC}(r_i) - \log\Sigma(r_i))^2}{(\delta \log n_{\rm GC}(r_i))^2} \cdot \frac{1}{DOF},
\end{equation}

where $n_{\rm GC}(r_i)$ is the number density of globular clusters at the $i$-th radius $r_i$, $\Sigma(r_i)$ is the stellar mass density at the same radius, $\delta$ denotes the uncertainty in logarithmic number density, and $DOF$ is the degrees of freedom, that is, the number of GC data points $-$ 1. We neglect the uncertainty in stellar mass because it is dominated by photometric zeropoint errors in the overlapping region. It corresponds only to a global shift, which we fit anyway by matching the profiles. The reduced $\chi^2$ equals 1 when the agreement is perfect and the uncertainties are estimated correctly.

The $\chi^2$ test is insensitive to different profile slopes as long as they agree within the error bars of the data points. Therefore, we also compare the effective radii $r_{\rm e}$ of de Vaucouleurs \citep{DeVaucouleurs1969} profile fits the data (grey dashed lines in Figure \ref{fig:fits}). For the GCs, we calculate the half-number radius $r_{\rm e,n}$ and for the mass profiles, we calculate the half-mass radius $r_{\rm e,\Sigma}$. The best-fit parameters are listed in Table \ref{tab:fitparams}. We note that the half-mass radii do not refer to the real mass distributions because the de Vaucouleurs profiles are fitted in a limited radial range. They should be interpreted as a measure of the local slope in the overlapping region.

\begin{table*}
  \centering
  \begin{tabular}{l|c|c|c|c}
     Component & Range (kpc) & $r_{\rm e,n}$ or $r_{\rm e,\Sigma}$ (kpc) & $n$ & $r_{\rm core}$ \\
     \hline \hline
     \textit{S\'ersic Fit} & & &\\
     ~~~Stellar Mass & full     & $2.875\pm0.076$ & $5.96\pm0.16$ & \ldots        \\
     ~~~Red GCs      & $r<40$   & $9.61\pm0.99$   & $3.10\pm0.80$ & \ldots        \\
     ~~~Blue GCs     & full     & $63\pm24$       & $4.17\pm0.93$ & \ldots        \\
     \hline
     \textit{de Vaucouleurs Fit} & & & \\
     ~~~Stellar Mass & $7<r<40$ & $6.80\pm0.39$   & 4             & \ldots        \\
     ~~~Red GCs      & $7<r<40$ & $7.6\pm1.4$     & 4             & \ldots        \\
     ~~~Total Mass (beta) & $2<r<60$ & $169\pm28$ & 4            & \ldots        \\
     ~~~Total Mass (NFW)  & $2<r<60$ & $17.6\pm1.8$ & 4             & \ldots        \\
     ~~~Blue GCs     & $2<r<60$ & $58.7\pm6.0$    & 4             & \ldots        \\
     \hline
     \textit{NFW Fit} & & & \\
     ~~~Total Mass   & full     & \ldots          & \ldots        & $1.53\pm0.12$ \\
     ~~~Red GCs      & full     & \ldots          & \ldots        & $2.42\pm0.91$ \\
     ~~~Blue GCs     & full     & \ldots          & \ldots        & $8.66\pm0.81$ \\
     \hline
  \end{tabular}
  \caption{Best-fit parameters to the projected mass density profiles (stellar and total mass) as well as to the number density profiles (GCs). The fitting range is given in column (2). Red GCs beyond $r>40$\,kpc are ignored due to high background contamination \citep{Usher2013}. The remaining columns are: (3) effective radii $r_{\rm e,\Sigma}$ and $r_{\rm e,n}$ for the mass and number density profiles, respectively, (4) S\'ersic index $n$, and (5) core radius $r_{\rm core}$ for the NFW fit. \label{tab:fitparams}}
\end{table*}

\subsection{Blue GCs trace Total Mass}

The blue GCs trace the total mass assuming it follows a mixture of a $\beta$ and an NFW profile. Given the large uncertainty of the model choice, the reduced $\chi^2=0.25$ is small. However, the reduced $\chi^2=1.6$ is much larger for the red GCs, meaning the blue GCs trace the total mass profile better than the red GCs.

Inside $r<15$\,kpc, the agreement is best between the blue GCs and the total mass assuming a $\beta$ model. Beyond $r>15$\,kpc, the $\beta$ model is too shallow while the NFW model is too steep (see Figure \ref{fig:massprofiles}). At these radii, the signal in the Chandra X-ray data is too faint to constrain the total mass profile well. If the blue GCs still trace it here, then the mass density slope must lie between $2<\gamma<3$ for $\rho\propto R^\gamma$. A deviation from the $\beta$ model at large radii to $\gamma>2$ is physically required because the $\beta$ profile has $\gamma=2$ and, hence, predicts diverging total mass.

The effective radius from a de Vaucouleurs fit to the blue GCs is $r_{\rm e,n}=58.7\pm6.0$\,kpc. Consistently, it is in between the effective radii of the $\beta$ profile ($r_{\rm e,\Sigma}=169\pm28$\,kpc) and the NFW profile ($r_{\rm e,\Sigma}=17.6\pm1.8$\,kpc).

The circular velocity $v_{\rm circ}$ of the total mass (Figure \ref{fig:gc_v_sigma}, continuous black line in the bottom panel) agrees with the velocity dispersion of the blue GCs (blue line) inside $r\lesssim10$\,kpc for the $\beta$ model. The $v_{\rm circ}$ is constant in this case because the $\beta$ model is isothermal. Beyond $r\gtrsim 6$\,kpc, the blue GC velocity dispersion falls below $v_{\rm circ}$ of the total mass, implying anisothermality. For the NFW model, the circular velocity is too high because more mass is enclosed inside $r\lesssim 10$\,kpc. 

The blue GC number density profile is well fit by a projected NFW profile \citep{Navarro1996} (see Figure \ref{fig:fits}, left panel) with a core radius of $r_{\rm core}=8.66\pm0.81$\,kpc. For the total mass profile assuming an NFW profile, the core radius is much smaller with $r_{\rm core}=1.53\pm0.12$\,kpc. This means, the transition from the shallower $\rho\propto R^{-1}$ to the steeper $\rho\propto R^{-3}$ density slope occurs farther in.

\subsection{Red GCs trace Stellar Mass} \label{sec:redtracestellarmass}

The red GCs trace the stellar mass between $7<r<40$\,kpc (see Figure \ref{fig:massprofiles}). In this region, we calculate a reduced $\chi^2=0.77$. This is significantly better than the reduced $\chi^2=35$ for the blue GCs and the stellar mass. At larger radii, the red GC slope becomes flatter, possibly due to increasing relative contamination by other sources \citep{Usher2013}. Using spectroscopic information, \cite{Usher2013} estimate 20\% contamination fraction in the Subaru sample at $r\approx25$\,kpc. It increases to 50\% at $r\approx40$\,kpc. A contamination correction would shift the GC data points downward by 0.1\,dex (0.3\,dex) at $r=20$\,kpc ($r=40$\,kpc). Inside of $r<20$\,kpc, the effect of contamination is negligible.

The effective radii of de Vaucouleurs function fits to the profiles, as shown in Figure \ref{fig:fits}, in the same radial region agree for the stellar mass ($r_{\rm e,\Sigma}=6.80\pm0.39$\,kpc) and the red GCs ($r_{\rm e,n}=7.6\pm1.4$\,kpc). The blue GC distribution is significantly more extended with a half-number radius $r_{\rm e,n}=58.7\pm6.0$\,kpc.

In addition to the resemblance of the 1-D surface mass and number density profiles, we find that the ellipticity and centering agree for both red and blue GCs subpopulations with the stellar mass within our $<2\sigma$ confidence intervals (see Figures \ref{fig:gc_distribution_sub} and \ref{fig:gc_isophotes}).

Since the red GCs trace the stellar mass, we would naturally expect both components to have similar velocity dispersion profiles. That is not the case. While they agree in the center, the stellar velocity dispersion profile (purple shade) decreases more steeply than the red GCs dispersion profile. In the absence of rotation (see Section \ref{sec:stellarkinematics}) to compensate for the declining dispersion, consequently, stars and red GCs must have different orbital anisotropy. A higher radial than tangential velocity dispersion can project to a smaller line-of-sight velocity dispersion at large radii for the same mass distribution \citep{Binney2008}.

\section{Discussion and Conclusions}

We have obtained new, deep, wide-field observations of the early-type galaxy NGC~4278 with the Wendelstein 43\,cm telescope. The $g$- and $i$-band observations are complemented by high-resolution archival Hubble Space Telescope imaging data in the F475W and F850LP bands.

The morphology of NGC~4278 is relaxed beyond $r>2$\,kpc. We find no signs of recent accretion events or interactions with the nearby galaxy NGC~4283. The presence of dust in the inner $r<2$\,kpc is aligned with a stellar rotation of $v\approx50$\,km\,s$^{-1}$, which suggests a past massive merger event \citep{hoffman2010} not more than 3\,Gyr ago \citep{schulze2017}. As this is limited to the very central part of the galaxy, it does not affect the morphology and kinematics at larger radii, which are of interest in this work.

We measure azimuthally averaged surface brightness profiles on all four images and combine them to obtain one stellar mass density profile. This profile is compared to the number density profiles of the blue and red globular clusters separately, which we adopt from \cite{Usher2013}. As found in other galaxies, the red GCs have a steeper radial profile than the blue ones. We discover that the red GCs trace the stellar mass density well out to very large radii, in the radial region $7<r<40$\,kpc, which corresponds to 2.4--14 stellar half-mass radii of NGC~4278.

Apart from the inner $r<2$\,kpc, there is no significant rotation of either the stellar light or the GCs. The small stellar ellipticity of $\epsilon\approx 0.1$ is consistent with both red and blue GCs. Moreover, both GC subpopulations are well centered on NGC 4278.

We also perform a new analysis of the total mass density profile using archival Chandra X-ray data and by assuming both a $\beta$ model and an NFW model. We find that the total mass density is well traced by the blue GCs within $r<15$\,kpc (5 stellar half-mass radii) for the $\beta$ model. Beyond $r>15$\,kpc, the blue GC number density falls steeper and is better traced by a mixture of the $\beta$ and NFW models. Unfortunately, at these large radii the X-ray signal eventually becomes too weak to further constrain the radial dark matter profile shape. Nevertheless, if the blue GCs continue to trace the total mass at these large radii, it implies the dark matter density $\rho$ depends on the 3-D radius $R$ via $\rho\propto R^\gamma$ with $2<\gamma<3$. This is in excellent agreement with the radial behaviour of the total density slope obtained from dynamical modelling \citep[e.g.,][]{Poci2017,Bellstedt2018} or strong lensing observations \citep[e.g.,][]{auger2010} for other early-type galaxies, and predicted from theory \citep{remus2013,remus2017}.

The physical implication of our results is that the formation of red GCs is closely connected to the stellar mass build-up of galaxies, while the blue GCs are mostly accreted together with low-mass galaxies in a more minor-merger oriented scenario. Red GCs are formed during massive gas-rich merger events. As such, they are either formed in the main galaxy itself or, alternatively, are partially brought in through a major merger event by the other massive galaxy that had previously formed many red GCs itself through similar gas-rich merger events. It is not possible to distinguish the red GCs that were formed by the galaxy itself from those that could have been accreted through the major merger, as such massive mergers usually mix both galaxies completely. Nevertheless, as the red GCs were always formed in more massive galaxies in-situ, the spatial distribution of the stellar light component and the red GCs evolve in unison because they were formed together. An example of this is NGC 1277, which is considered a red nugget, and lacks the second phase of galaxy formation with associated blue GCs \citep{Beasley2018}, and in fact is dominated by red GCs.

On the other hand, blue GCs are brought in mostly through minor and mini mergers. According to the model by \cite{Valenzuela2021}, subhalos with a total mass of $M_{\rm seedGC} \approx 4\times10^8$\,M$_\odot$ form on average one GC in parallel to their stars. As these subhalos merge with others, the number of blue GCs increases hierarchically in parallel to the total mass of the galaxy. In particular, as the galaxies of such small mass merger events usually deposit most of their mass at large radii \citep[e.g.,][]{karademir2019}, blue GCs are also deposited more commonly at larger radii, and more closely bound to the overall potential than tracing the stellar component of the main galaxy. This is in excellent agreement with our result that the blue GCs trace the underlying total mass distribution and not only the light component of the galaxy. \textit{Therefore, we conclude that blue GCs are excellent tracers of the total, dark matter dominated underlying potential of a galaxy, and thus can be used to trace the dark matter.} This is particularly useful for galaxies where no X-ray component can be measured.

\section*{Acknowledgements}

We wish to thank the anonymous referee for his or her comments and suggestions that allowed us to improve the paper.

Iu.B. acknowledges support from XMM-Newton Grant Number
80NSSC19K1056. The 40cm telescope was funded by Ludwig-Maximilians-University, Munich. Some of the upgrades for the infrastructure and the 40cm telescope housing were funded by the Cluster of Excellence ``Origin of the Universe" of the German Science Foundation DFG.

This work made use of data products based on observations made with the NASA/ESA \textit{Hubble Space Telescope} and obtained from the Hubble Legacy Archive, which is a collaboration between the Space Telescope Science Institute (STScI/NASA), the Space Telescope European Coordinating Facility (ST-ECF/ESA), and the Canadian Astronomy Data Centre (CADC/NRC/CSA).

We also used the image display tool SAOImage DS9 developed by Smithsonian Astrophysical Observatory and the image display tool Fitsedit, developed by Johannes Koppenhoefer.

\section*{Data Availability}

Some data used in this project are taken from published works of \cite{Usher2013} (GC spatial distribution), \cite{forbes2017b} (GC kinematics), and the Hubble Legacy Archive (HST images of NGC 4278). Reasonable requests for data can be made to the corresponding author.

\bibliographystyle{mnras}
\bibliography{ngc4278}{}

\begin{thebibliography}{}
\makeatletter
\relax
\def\mn@urlcharsother{\let\do\@makeother \do\$\do\&\do\#\do\^\do\_\do\%\do\~}
\def\mn@doi{\begingroup\mn@urlcharsother \@ifnextchar [ {\mn@doi@}
  {\mn@doi@[]}}
\def\mn@doi@[#1]#2{\def\@tempa{#1}\ifx\@tempa\@empty \href
  {http://dx.doi.org/#2} {doi:#2}\else \href {http://dx.doi.org/#2} {#1}\fi
  \endgroup}
\def\mn@eprint#1#2{\mn@eprint@#1:#2::\@nil}
\def\mn@eprint@arXiv#1{\href {http://arxiv.org/abs/#1} {{\tt arXiv:#1}}}
\def\mn@eprint@dblp#1{\href {http://dblp.uni-trier.de/rec/bibtex/#1.xml}
  {dblp:#1}}
\def\mn@eprint@#1:#2:#3:#4\@nil{\def\@tempa {#1}\def\@tempb {#2}\def\@tempc
  {#3}\ifx \@tempc \@empty \let \@tempc \@tempb \let \@tempb \@tempa \fi \ifx
  \@tempb \@empty \def\@tempb {arXiv}\fi \@ifundefined
  {mn@eprint@\@tempb}{\@tempb:\@tempc}{\expandafter \expandafter \csname
  mn@eprint@\@tempb\endcsname \expandafter{\@tempc}}}

\bibitem[\protect\citeauthoryear{{Auger}, {Treu}, {Bolton}, {Gavazzi},
  {Koopmans}, {Marshall}, {Moustakas}  \& {Burles}}{{Auger}
  et~al.}{2010}]{auger2010}
{Auger} M.~W.,  {Treu} T.,  {Bolton} A.~S.,  {Gavazzi} R.,  {Koopmans}
  L.~V.~E.,  {Marshall} P.~J.,  {Moustakas} L.~A.,   {Burles} S.,  2010,
  \mn@doi [\apj] {10.1088/0004-637X/724/1/511}, \href
  {https://ui.adsabs.harvard.edu/abs/2010ApJ...724..511A} {724, 511}

\bibitem[\protect\citeauthoryear{{Babyk}, {McNamara}, {Nulsen}, {Hogan},
  {Vantyghem}, {Russell}, {Pulido}  \& {Edge}}{{Babyk}
  et~al.}{2018}]{Babyk2018}
{Babyk} I.~V.,  {McNamara} B.~R.,  {Nulsen} P.~E.~J.,  {Hogan} M.~T.,
  {Vantyghem} A.~N.,  {Russell} H.~R.,  {Pulido} F.~A.,   {Edge} A.~C.,  2018,
  \mn@doi [\apj] {10.3847/1538-4357/aab3c9}, \href
  {https://ui.adsabs.harvard.edu/abs/2018ApJ...857...32B} {857, 32}

\bibitem[\protect\citeauthoryear{{Beasley}, {Baugh}, {Forbes}, {Sharples}  \&
  {Frenk}}{{Beasley} et~al.}{2002}]{beasley2002}
{Beasley} M.~A.,  {Baugh} C.~M.,  {Forbes} D.~A.,  {Sharples} R.~M.,   {Frenk}
  C.~S.,  2002, \mn@doi [\mnras] {10.1046/j.1365-8711.2002.05402.x}, \href
  {https://ui.adsabs.harvard.edu/abs/2002MNRAS.333..383B} {333, 383}

\bibitem[\protect\citeauthoryear{{Beasley}, {Trujillo}, {Leaman}  \&
  {Montes}}{{Beasley} et~al.}{2018}]{Beasley2018}
{Beasley} M.~A.,  {Trujillo} I.,  {Leaman} R.,   {Montes} M.,  2018, \mn@doi
  [\nat] {10.1038/nature25756}, \href
  {https://ui.adsabs.harvard.edu/abs/2018Natur.555..483B} {555, 483}

\bibitem[\protect\citeauthoryear{{Bellstedt} et~al.,}{{Bellstedt}
  et~al.}{2018}]{Bellstedt2018}
{Bellstedt} S.,  et~al., 2018, \mn@doi [\mnras] {10.1093/mnras/sty456}, \href
  {https://ui.adsabs.harvard.edu/abs/2018MNRAS.476.4543B} {476, 4543}

\bibitem[\protect\citeauthoryear{{Binney} \& {Tremaine}}{{Binney} \&
  {Tremaine}}{2008}]{Binney2008}
{Binney} J.,  {Tremaine} S.,  2008, {Galactic Dynamics: Second Edition}.
Princeton, NJ, Princeton University Press

\bibitem[\protect\citeauthoryear{{Blakeslee}, {Tonry}  \&
  {Metzger}}{{Blakeslee} et~al.}{1997}]{blakeslee1997}
{Blakeslee} J.~P.,  {Tonry} J.~L.,   {Metzger} M.~R.,  1997, \mn@doi [\aj]
  {10.1086/118488}, \href
  {https://ui.adsabs.harvard.edu/abs/1997AJ....114..482B} {114, 482}

\bibitem[\protect\citeauthoryear{{Bogd{\'a}n}, {David}, {Jones}, {Forman}  \&
  {Kraft}}{{Bogd{\'a}n} et~al.}{2012}]{bodgan2012}
{Bogd{\'a}n} {\'A}.,  {David} L.~P.,  {Jones} C.,  {Forman} W.~R.,   {Kraft}
  R.~P.,  2012, \mn@doi [\apj] {10.1088/0004-637X/758/1/65}, \href
  {https://ui.adsabs.harvard.edu/abs/2012ApJ...758...65B} {758, 65}

\bibitem[\protect\citeauthoryear{{Bradley} et~al.,}{{Bradley}
  et~al.}{2021}]{Bradley2021}
{Bradley} L.,  et~al., 2021, {astropy/photutils: 1.1.0}, Zenodo,
  \mn@doi{10.5281/zenodo.4624996}

\bibitem[\protect\citeauthoryear{{Brodie} \& {Strader}}{{Brodie} \&
  {Strader}}{2006}]{Brodie2006}
{Brodie} J.~P.,  {Strader} J.,  2006, \mn@doi [\araa]
  {10.1146/annurev.astro.44.051905.092441}, \href
  {https://ui.adsabs.harvard.edu/abs/2006ARA&A..44..193B} {44, 193}

\bibitem[\protect\citeauthoryear{{Brodie} et~al.,}{{Brodie}
  et~al.}{2014}]{Brodie2014}
{Brodie} J.~P.,  et~al., 2014, \mn@doi [\apj] {10.1088/0004-637X/796/1/52},
  \href {https://ui.adsabs.harvard.edu/abs/2014ApJ...796...52B} {796, 52}

\bibitem[\protect\citeauthoryear{{Burkert} \& {Forbes}}{{Burkert} \&
  {Forbes}}{2020}]{Burkert2020}
{Burkert} A.,  {Forbes} D.~A.,  2020, \mn@doi [\aj] {10.3847/1538-3881/ab5b0e},
  \href {https://ui.adsabs.harvard.edu/abs/2020AJ....159...56B} {159, 56}

\bibitem[\protect\citeauthoryear{{Cavaliere} \& {Fusco-Femiano}}{{Cavaliere} \&
  {Fusco-Femiano}}{1978}]{Cavaliere1978}
{Cavaliere} A.,  {Fusco-Femiano} R.,  1978, \aap, \href
  {https://ui.adsabs.harvard.edu/abs/1978A&A....70..677C} {70, 677}

\bibitem[\protect\citeauthoryear{{D'Abrusco}, {Fabbiano}  \&
  {Brassington}}{{D'Abrusco} et~al.}{2014}]{dabrusco2014}
{D'Abrusco} R.,  {Fabbiano} G.,   {Brassington} N.~J.,  2014, \mn@doi [\apj]
  {10.1088/0004-637X/783/1/19}, \href
  {https://ui.adsabs.harvard.edu/abs/2014ApJ...783...19D} {783, 19}

\bibitem[\protect\citeauthoryear{{Dolfi}, {Forbes}, {Couch}, {Bekki},
  {Ferr{\'e}-Mateu}, {Romanowsky}  \& {Brodie}}{{Dolfi}
  et~al.}{2021}]{dolfi2021}
{Dolfi} A.,  {Forbes} D.~A.,  {Couch} W.~J.,  {Bekki} K.,  {Ferr{\'e}-Mateu}
  A.,  {Romanowsky} A.~J.,   {Brodie} J.~P.,  2021, \mn@doi [\mnras]
  {10.1093/mnras/stab1023}, \href
  {https://ui.adsabs.harvard.edu/abs/2021MNRAS.504.4923D} {504, 4923}

\bibitem[\protect\citeauthoryear{{Forbes} \& {Remus}}{{Forbes} \&
  {Remus}}{2018}]{Forbes2018b}
{Forbes} D.~A.,  {Remus} R.-S.,  2018, \mn@doi [\mnras]
  {10.1093/mnras/sty1767}, \href
  {https://ui.adsabs.harvard.edu/abs/2018MNRAS.479.4760F} {479, 4760}

\bibitem[\protect\citeauthoryear{{Forbes}, {Ponman}  \& {O'Sullivan}}{{Forbes}
  et~al.}{2012}]{Forbes2012}
{Forbes} D.~A.,  {Ponman} T.,   {O'Sullivan} E.,  2012, \mn@doi [\mnras]
  {10.1111/j.1365-2966.2012.21368.x}, \href
  {https://ui.adsabs.harvard.edu/abs/2012MNRAS.425...66F} {425, 66}

\bibitem[\protect\citeauthoryear{{Forbes} et~al.,}{{Forbes}
  et~al.}{2017a}]{forbes2017c}
{Forbes} D.~A.,  et~al., 2017a, \mn@doi [\aj] {10.3847/1538-3881/153/3/114},
  \href {https://ui.adsabs.harvard.edu/abs/2017AJ....153..114F} {153, 114}

\bibitem[\protect\citeauthoryear{{Forbes}, {Sinpetru}, {Savorgnan},
  {Romanowsky}, {Usher}  \& {Brodie}}{{Forbes} et~al.}{2017b}]{forbes2017b}
{Forbes} D.~A.,  {Sinpetru} L.,  {Savorgnan} G.,  {Romanowsky} A.~J.,  {Usher}
  C.,   {Brodie} J.,  2017b, \mn@doi [\mnras] {10.1093/mnras/stw2604}, \href
  {https://ui.adsabs.harvard.edu/abs/2017MNRAS.464.4611F} {464, 4611}

\bibitem[\protect\citeauthoryear{{Forbes} et~al.,}{{Forbes}
  et~al.}{2018a}]{Forbes2018}
{Forbes} D.~A.,  et~al., 2018a, \mn@doi [Proceedings of the Royal Society of
  London Series A] {10.1098/rspa.2017.0616}, \href
  {https://ui.adsabs.harvard.edu/abs/2018RSPSA.47470616F} {474, 20170616}

\bibitem[\protect\citeauthoryear{{Forbes}, {Read}, {Gieles}  \&
  {Collins}}{{Forbes} et~al.}{2018b}]{forbes2018c}
{Forbes} D.~A.,  {Read} J.~I.,  {Gieles} M.,   {Collins} M. L.~M.,  2018b,
  \mn@doi [\mnras] {10.1093/mnras/sty2584}, \href
  {https://ui.adsabs.harvard.edu/abs/2018MNRAS.481.5592F} {481, 5592}

\bibitem[\protect\citeauthoryear{{Forte}, {Faifer}  \& {Geisler}}{{Forte}
  et~al.}{2005}]{Forte2005}
{Forte} J.~C.,  {Faifer} F.,   {Geisler} D.,  2005, \mn@doi [\mnras]
  {10.1111/j.1365-2966.2004.08572.x}, \href
  {https://ui.adsabs.harvard.edu/abs/2005MNRAS.357...56F} {357, 56}

\bibitem[\protect\citeauthoryear{{Foster} et~al.,}{{Foster}
  et~al.}{2016}]{foster2016}
{Foster} C.,  et~al., 2016, \mn@doi [\mnras] {10.1093/mnras/stv2947}, \href
  {https://ui.adsabs.harvard.edu/abs/2016MNRAS.457..147F} {457, 147}

\bibitem[\protect\citeauthoryear{{Harris}, {Harris}  \& {Hudson}}{{Harris}
  et~al.}{2015}]{Harris2015}
{Harris} W.~E.,  {Harris} G.~L.,   {Hudson} M.~J.,  2015, \mn@doi [\apj]
  {10.1088/0004-637X/806/1/36}, \href
  {https://ui.adsabs.harvard.edu/abs/2015ApJ...806...36H} {806, 36}

\bibitem[\protect\citeauthoryear{{Hoffman}, {Cox}, {Dutta}  \&
  {Hernquist}}{{Hoffman} et~al.}{2010}]{hoffman2010}
{Hoffman} L.,  {Cox} T.~J.,  {Dutta} S.,   {Hernquist} L.,  2010, \mn@doi
  [\apj] {10.1088/0004-637X/723/1/818}, \href
  {https://ui.adsabs.harvard.edu/abs/2010ApJ...723..818H} {723, 818}

\bibitem[\protect\citeauthoryear{{Karademir}, {Remus}, {Burkert}, {Dolag},
  {Hoffmann}, {Moster}, {Steinwandel}  \& {Zhang}}{{Karademir}
  et~al.}{2019}]{karademir2019}
{Karademir} G.~S.,  {Remus} R.-S.,  {Burkert} A.,  {Dolag} K.,  {Hoffmann}
  T.~L.,  {Moster} B.~P.,  {Steinwandel} U.~P.,   {Zhang} J.,  2019, \mn@doi
  [\mnras] {10.1093/mnras/stz1251}, \href
  {https://ui.adsabs.harvard.edu/abs/2019MNRAS.487..318K} {487, 318}

\bibitem[\protect\citeauthoryear{{Kluge} et~al.,}{{Kluge}
  et~al.}{2020}]{Kluge2020}
{Kluge} M.,  et~al., 2020, \mn@doi [\apjs] {10.3847/1538-4365/ab733b}, \href
  {https://ui.adsabs.harvard.edu/abs/2020ApJS..247...43K} {247, 43}

\bibitem[\protect\citeauthoryear{{Li} \& {Gnedin}}{{Li} \&
  {Gnedin}}{2014}]{2014ApJ...796...10L}
{Li} H.,  {Gnedin} O.~Y.,  2014, \mn@doi [\apj] {10.1088/0004-637X/796/1/10},
  \href {https://ui.adsabs.harvard.edu/abs/2014ApJ...796...10L} {796, 10}

\bibitem[\protect\citeauthoryear{{Navarro}, {Frenk}  \& {White}}{{Navarro}
  et~al.}{1996}]{Navarro1996}
{Navarro} J.~F.,  {Frenk} C.~S.,   {White} S. D.~M.,  1996, \mn@doi [\apj]
  {10.1086/177173}, \href
  {https://ui.adsabs.harvard.edu/abs/1996ApJ...462..563N} {462, 563}

\bibitem[\protect\citeauthoryear{{Pellegrini}, {Wang}, {Fabbiano}, {Kim},
  {Brassington}, {Gallagher}, {Trinchieri}  \& {Zezas}}{{Pellegrini}
  et~al.}{2012}]{pellegrini2012}
{Pellegrini} S.,  {Wang} J.,  {Fabbiano} G.,  {Kim} D.-W.,  {Brassington}
  N.~J.,  {Gallagher} J.~S.,  {Trinchieri} G.,   {Zezas} A.,  2012, \mn@doi
  [\apj] {10.1088/0004-637X/758/2/94}, \href
  {https://ui.adsabs.harvard.edu/abs/2012ApJ...758...94P} {758, 94}

\bibitem[\protect\citeauthoryear{{Peng}, {Ford}  \& {Freeman}}{{Peng}
  et~al.}{2004}]{peng2004}
{Peng} E.~W.,  {Ford} H.~C.,   {Freeman} K.~C.,  2004, \mn@doi [\apj]
  {10.1086/381236}, \href
  {https://ui.adsabs.harvard.edu/abs/2004ApJ...602..705P} {602, 705}

\bibitem[\protect\citeauthoryear{{Poci}, {Cappellari}  \& {McDermid}}{{Poci}
  et~al.}{2017}]{Poci2017}
{Poci} A.,  {Cappellari} M.,   {McDermid} R.~M.,  2017, \mn@doi [\mnras]
  {10.1093/mnras/stx101}, \href
  {https://ui.adsabs.harvard.edu/abs/2017MNRAS.467.1397P} {467, 1397}

\bibitem[\protect\citeauthoryear{{Pota} et~al.,}{{Pota}
  et~al.}{2013}]{pota2013}
{Pota} V.,  et~al., 2013, \mn@doi [\mnras] {10.1093/mnras/sts029}, \href
  {https://ui.adsabs.harvard.edu/abs/2013MNRAS.428..389P} {428, 389}

\bibitem[\protect\citeauthoryear{{Reina-Campos}, {Trujillo-Gomez}, {Pfeffer},
  {Sills}, {Deason}, {Crain}  \& {Kruijssen}}{{Reina-Campos}
  et~al.}{2022a}]{Reina-Campos2022a}
{Reina-Campos} M.,  {Trujillo-Gomez} S.,  {Pfeffer} J.~L.,  {Sills} A.,
  {Deason} A.~J.,  {Crain} R.~A.,   {Kruijssen} J.~M.~D.,  2022a, arXiv
  e-prints, \href {https://ui.adsabs.harvard.edu/abs/2022arXiv220411861R} {p.
  arXiv:2204.11861}

\bibitem[\protect\citeauthoryear{{Reina-Campos}, {Trujillo-Gomez}, {Deason},
  {Kruijssen}, {Pfeffer}, {Crain}, {Bastian}  \& {Hughes}}{{Reina-Campos}
  et~al.}{2022b}]{Reina-Campos2022b}
{Reina-Campos} M.,  {Trujillo-Gomez} S.,  {Deason} A.~J.,  {Kruijssen}
  J.~M.~D.,  {Pfeffer} J.~L.,  {Crain} R.~A.,  {Bastian} N.,   {Hughes} M.~E.,
  2022b, \mn@doi [\mnras] {10.1093/mnras/stac1126}, \href
  {https://ui.adsabs.harvard.edu/abs/2022MNRAS.513.3925R} {513, 3925}

\bibitem[\protect\citeauthoryear{{Remus}, {Burkert}, {Dolag}, {Johansson},
  {Naab}, {Oser}  \& {Thomas}}{{Remus} et~al.}{2013}]{remus2013}
{Remus} R.-S.,  {Burkert} A.,  {Dolag} K.,  {Johansson} P.~H.,  {Naab} T.,
  {Oser} L.,   {Thomas} J.,  2013, \mn@doi [\apj] {10.1088/0004-637X/766/2/71},
  \href {https://ui.adsabs.harvard.edu/abs/2013ApJ...766...71R} {766, 71}

\bibitem[\protect\citeauthoryear{{Remus}, {Dolag}, {Naab}, {Burkert},
  {Hirschmann}, {Hoffmann}  \& {Johansson}}{{Remus} et~al.}{2017}]{remus2017}
{Remus} R.-S.,  {Dolag} K.,  {Naab} T.,  {Burkert} A.,  {Hirschmann} M.,
  {Hoffmann} T.~L.,   {Johansson} P.~H.,  2017, \mn@doi [\mnras]
  {10.1093/mnras/stw2594}, \href
  {https://ui.adsabs.harvard.edu/abs/2017MNRAS.464.3742R} {464, 3742}

\bibitem[\protect\citeauthoryear{{Roediger} \& {Courteau}}{{Roediger} \&
  {Courteau}}{2015}]{Roediger2015}
{Roediger} J.~C.,  {Courteau} S.,  2015, \mn@doi [\mnras]
  {10.1093/mnras/stv1499}, \href
  {https://ui.adsabs.harvard.edu/abs/2015MNRAS.452.3209R} {452, 3209}

\bibitem[\protect\citeauthoryear{{Schlafly} \& {Finkbeiner}}{{Schlafly} \&
  {Finkbeiner}}{2011}]{Schlafly2011}
{Schlafly} E.~F.,  {Finkbeiner} D.~P.,  2011, \mn@doi [\apj]
  {10.1088/0004-637X/737/2/103}, \href
  {http://adsabs.harvard.edu/abs/2011ApJ...737..103S} {737, 103}

\bibitem[\protect\citeauthoryear{{Schuberth}, {Richtler}, {Hilker}, {Dirsch},
  {Bassino}, {Romanowsky}  \& {Infante}}{{Schuberth}
  et~al.}{2010}]{schuberth2010}
{Schuberth} Y.,  {Richtler} T.,  {Hilker} M.,  {Dirsch} B.,  {Bassino} L.~P.,
  {Romanowsky} A.~J.,   {Infante} L.,  2010, \mn@doi [\aap]
  {10.1051/0004-6361/200912482}, \href
  {https://ui.adsabs.harvard.edu/abs/2010A&A...513A..52S} {513, A52}

\bibitem[\protect\citeauthoryear{{Schulze}, {Remus}  \& {Dolag}}{{Schulze}
  et~al.}{2017}]{schulze2017}
{Schulze} F.,  {Remus} R.-S.,   {Dolag} K.,  2017, \mn@doi [Galaxies]
  {10.3390/galaxies5030041}, \href
  {https://ui.adsabs.harvard.edu/abs/2017Galax...5...41S} {5, 41}

\bibitem[\protect\citeauthoryear{{S\'ersic}}{{S\'ersic}}{1968}]{Sersic1968}
{S\'ersic} J.~L.,  1968, {Atlas de Galaxias Australes (C\'ordoba: Observatorio
  Astronomico, Univ. C\'ordoba)}

\bibitem[\protect\citeauthoryear{{Spitler} \& {Forbes}}{{Spitler} \&
  {Forbes}}{2009}]{Spitler2009}
{Spitler} L.~R.,  {Forbes} D.~A.,  2009, \mn@doi [\mnras]
  {10.1111/j.1745-3933.2008.00567.x}, \href
  {https://ui.adsabs.harvard.edu/abs/2009MNRAS.392L...1S} {392, L1}

\bibitem[\protect\citeauthoryear{{Strader} et~al.,}{{Strader}
  et~al.}{2011}]{Strader2011}
{Strader} J.,  et~al., 2011, \mn@doi [\apjs] {10.1088/0067-0049/197/2/33},
  \href {https://ui.adsabs.harvard.edu/abs/2011ApJS..197...33S} {197, 33}

\bibitem[\protect\citeauthoryear{{Tully} et~al.,}{{Tully}
  et~al.}{2013}]{Tully2013}
{Tully} R.~B.,  et~al., 2013, \mn@doi [\aj] {10.1088/0004-6256/146/4/86}, \href
  {https://ui.adsabs.harvard.edu/abs/2013AJ....146...86T} {146, 86}

\bibitem[\protect\citeauthoryear{{Usher} et~al.,}{{Usher}
  et~al.}{2012}]{Usher2012}
{Usher} C.,  et~al., 2012, \mn@doi [\mnras] {10.1111/j.1365-2966.2012.21801.x},
  \href {https://ui.adsabs.harvard.edu/abs/2012MNRAS.426.1475U} {426, 1475}

\bibitem[\protect\citeauthoryear{{Usher}, {Forbes}, {Spitler}, {Brodie},
  {Romanowsky}, {Strader}  \& {Woodley}}{{Usher} et~al.}{2013}]{Usher2013}
{Usher} C.,  {Forbes} D.~A.,  {Spitler} L.~R.,  {Brodie} J.~P.,  {Romanowsky}
  A.~J.,  {Strader} J.,   {Woodley} K.~A.,  2013, \mn@doi [\mnras]
  {10.1093/mnras/stt1637}, \href
  {https://ui.adsabs.harvard.edu/abs/2013MNRAS.436.1172U} {436, 1172}

\bibitem[\protect\citeauthoryear{{Valenzuela}, {Moster}, {Remus}, {O'Leary}  \&
  {Burkert}}{{Valenzuela} et~al.}{2021}]{Valenzuela2021}
{Valenzuela} L.~M.,  {Moster} B.~P.,  {Remus} R.-S.,  {O'Leary} J.~A.,
  {Burkert} A.,  2021, \mn@doi [\mnras] {10.1093/mnras/stab1701}, \href
  {https://ui.adsabs.harvard.edu/abs/2021MNRAS.505.5815V} {505, 5815}

\bibitem[\protect\citeauthoryear{{de Vaucouleurs}}{{de
  Vaucouleurs}}{1969}]{DeVaucouleurs1969}
{de Vaucouleurs} G.,  1969, \aplett, \href
  {http://adsabs.harvard.edu/abs/1969ApL.....4...17D} {4, 17}

\makeatother
\end{thebibliography}

\bsp	
\label{lastpage}
\end{document}